\documentclass[prb,twocolumn,aps,noshowpacs]{revtex4}
\usepackage{graphicx,epsfig}
\usepackage{bm}
\usepackage{amssymb} 
\usepackage{bbold}

\begin{document}

\title{Slow oscillating dynamics of a two-level
system subject to a fast telegraph noise: beyond the NIBA
approximation}

\author{V. V. Mkhitaryan  and M. E. Raikh}

\affiliation{ Department of Physics and
Astronomy, University of Utah, Salt Lake City, UT 84112}

\begin{abstract}

We study the dynamics of a two-site model in which the tunneling
amplitude between the sites is not constant but rather a
high-frequency noise. Obviously, the population imbalance in this
model decays exponentially with time. Remarkably, the decay is
modified dramatically when the level asymmetry fluctuates {\em
in-phase} with fluctuations of the tunneling amplitude. For
particular type of these in-phase fluctuations,
 namely, the telegraph noise, we find the exact solution
for the average population dynamics. It appears that the
population imbalance  between the sites starting from $1$ at time
$t=0$ approaches a constant value in the limit $t\rightarrow
\infty$. At finite bias, the imbalance goes to zero at
$t\rightarrow \infty$, while the dynamics of the decay governed by
noise acquires an oscillatory character.

\end{abstract}

\maketitle

\section{Introduction}

The central problem in the field of dissipative
dynamics\cite{1,2,3,4,5,6,7,8,9} is formulated as follows.
Consider a two-site system described by the Hamiltonian
\begin{equation}
H= \frac 12\Big[\Delta \sigma_x + \varepsilon \sigma_z \Big ],
\end{equation}
where $\Delta$ denotes the tunneling amplitude between the sites,
while $\varepsilon$ is the detuning between the on-site levels,
i.e. the bias. Without interaction with bath, the dynamics of the
system contains a single frequency $\omega =\left(\Delta^2
+\varepsilon^2 \right)^{1/2}$. Dissipative dynamics studies how
the interaction with the bath in the form of the random modulation
of $\varepsilon$ slows down the oscillations of the population,
$\sigma_z(t)$, of the left site occupied at $t=0$.

Original results\cite{1} on dissipative dynamics were obtained
within non-interacting blip approximation (NIBA). The accuracy of
NIBA, see e.g. Refs. \onlinecite{C1,C2,C3,C4,C5,C6,C7,C8}, is
evaluated by comparison of the numerical results for bath-averaged
$\langle\sigma_z(t)\rangle$ obtained within NIBA with numerical
results obtained within different versions of the master
equations.

Certainly there is always a question: in what domain of the bath
parameters NIBA is applicable. A related question is: are there
physical effects which are not captured by NIBA.

In order to simplify the analysis as much as possible
the authors of Ref. \onlinecite{Imambekov} considered the case of
very high temperatures when the bath can be viewed as a fast
classical noise, $b_z(t)$. Then the equations of motion, $
\frac{d{\bf S}}{d t} ={\bf B}\times{\bf S}$, where ${\bf
B}=(\Delta, 0, b_z(t))$, i.e. at zero bias, can be reduced to the
following closed equation for $\sigma_z(t)$\cite{Dekker1987}
\begin{equation}
\label{Szeq}
\frac{d \sigma_z}{d t} =-\Delta^2\int
\limits_0^t dt_1 \cos\left[\int_{t_1}^t dt_2
b_z(t_2) \right ] \sigma_z(t_1),
\end{equation}
which applies for an arbitrary realization of the noise. We are
interested in the noise-averaged $\langle \sigma_z(t)\rangle$.
NIBA approach corresponds to the decoupling of average of the
product in the integrand of Eq. (\ref{Szeq}) into the product of averages
\begin{equation}
\langle \cos\left[\int_{t_1}^t dt_2
b_z(t_2) \right ] \sigma_z(t_1) \rangle=
\langle \cos\left[\int_{t_1}^t dt_2
b_z(t_2) \right ]\rangle \langle \sigma_z(t_1) \rangle.
\end{equation}
For the Gaussian white noise one has $\langle
\cos\left[\int_{t_1}^t dt_2 b_z(t_2) \right ]\rangle
=\exp[-\Gamma(t-t_1)]$. The magnitude, $b_z$, and the short
correlation time, $\tau_z$, are encoded into parameter $\Gamma\sim
b_z^2\tau_z$.
The integral equation Eq. (\ref{Szeq}) reduces to the second-order
differential equation  for the average $\langle \sigma_z(t)
\rangle$
\begin{equation}
\label{diffeq}
\frac{d^2 \langle \sigma_z \rangle}{d t^2}+ \Gamma
\frac{d \langle \sigma_z \rangle}{d t} +\Delta^2
\langle \sigma_z \rangle =0.
\end{equation}
It should be noted that the white-noise assumption,
$b_z\tau_z\ll1$, justifies NIBA. Indeed, the dynamics described by
Eq. (\ref{diffeq}) has two characteristic times, $1/\Gamma$ and
$\Gamma/\Delta^2\gg 1/\Gamma$. For white noise, these times are
much longer than $\tau_z$.

Assume now that, instead of a constant $\Delta$, we have some
random $b_x(t)$. This hypothetical situation implies that
tunneling between the sites, constituting a two-state system, is
exclusively due to noise. Then the  NIBA ansatz prescribes
two independent averagings in
the integrand. We thus get
\begin{equation}
\label{SzNIBA}
\frac{d \langle \sigma_z \rangle}{d t} =-\int
\limits_0^t dt_1\left \langle b_x(t) b_x(t_1) \cos\left[\int_{t_1}^t dt_2
b_z(t_2) \right ]\right \rangle \langle \sigma_z(t_1) \rangle,
\end{equation}
Upon performing extra averaging in the kernel with the help of
$\langle b_x(t)b_x(t_1)\rangle =b_x^2\exp[(t-t_1)/\tau_x]$, we
find that $\langle \sigma_z(t) \rangle$ exhibits a simple
exponential decay,
\begin{equation}
\label{Szpureexp}
\langle \sigma_z(t) \rangle= \exp\left [-\big (b_x^2\tau_x
+b_z^2\tau_z\big )t\right ],
\end{equation}
with a single characteristic time.
Again, NIBA is justified when both $b_x\tau_x$ and $b_z\tau_z$ are small.

The central question addressed in the present paper is: what
happens when the noise components, $b_x(t)$ and $b_z(t)$, being
both fast, are strongly correlated? What makes this question
non-trivial is the fact that the average of $b_x(t) b_x(t_1)
\cos\left[\int_{t_1}^t dt_2 b_z(t_2) \right ]$ now contains, in
addition to fast, a {\em slow contribution}. On the other hand,
for applicability of NIBA, this average should change faster than
$\langle \sigma_z(t)\rangle$. Thus we ask ourselves: what is the
spin dynamics when the condition of applicability of NIBA is
violated? Fortunately, the answer to this question can be obtained
purely analytically. This is because, for the telegraph noise, the
average spin dynamics can be found exactly. We obtain this result
in Sect. II. Comparing the NIBA and exact results, we demonstrate that NIBA
applies for $b_x\ll b_z$ . It appears that, unlike
Eq. (\ref{Szpureexp}), with correlated  $b_x(t)$ and $b_z(t)$, both
NIBA and exact results {\em saturate} at long time. In Sect. III we
establish that the saturation takes
place only at zero bias, $\varepsilon=0$. At any finite bias,
the average,  $\langle \sigma_z(t) \rangle$, decays with time.

\section{Zero bias}

\subsection{NIBA}

As it was demonstrated in Ref. \onlinecite{we} (see also the
Appendix), for $b_x(t)$ and $b_z(t)$ in the form of the telegraph
noise, the kernel in the NIBA equation Eq. (\ref{SzNIBA}) has the
following form
\begin{eqnarray}
\label{KT} K(T)\!\!&=&\!\!\left \langle b_x(t) b_x(t+T)
\cos\left[\int_{t}^{t+T} dt'
b_z(t') \right ]\right \rangle \\
&=&\!\!\frac{b_x^2} {\tau_s-\tau_f} \left[
\tau_s\exp\!\left(\!-\frac{T}{\tau_{f}}\right)
-\tau_f\exp\!\left(\!-\frac{T}{\tau_s}\right) \right], \nonumber
\end{eqnarray}
where $\tau_f$ and $\tau_s$ denote the fast and slow relaxation
times defined as
\begin{equation}
\label{taufs}
\tau_f=\frac{\tau}{1+
\big(1-b_z^2\tau^2\big)^{1/2}},\quad \tau_s=\frac{\tau}{1-
\big(1-b_z^2\tau^2\big)^{1/2}}.
\end{equation}
The NIBA equation with the kernel Eq. (\ref{KT}) yields
the solution
\begin{eqnarray}
\label{SzsolNIBA}
&&\langle \sigma_z(t)\rangle =\exp\Big[-\int_0^tdt_1\int_0^{t_1}dt_2 K(t_1-t_2)\Big]\\
&&=\exp\!\Bigg \{\!-\frac{b_x^2\tau_f\tau_s}{\tau_s-\tau_f}
\left[\tau_s\left(1- e^{-t/{\tau_s}}\right) -\tau_f \left(1-
e^{- t/{\tau_f}}\right)\right]\! \Bigg \}.\nonumber
\end{eqnarray}
At long times the solution Eq. (\ref{SzsolNIBA}) saturates at the
value
\begin{equation}
\label{larget}
\langle \sigma_z(t)\rangle\big\vert_{t\to\infty} =
 \exp\left(-b_x^2\tau_s\tau_f\right)=\exp\left(-\frac{b_x^2}{b_z^2}\right).
\end{equation}

\subsection{Exact solution}

The key observation, which allows to solve the
problem exactly, is that, with telegraph noise,
the eigenvectors of the time-dependent Hamiltonian
\begin{equation}
\label{allnoise} H= \frac 12\Big[ b_x(t) \sigma_x + b_z(t)\sigma_z
\Big]
\end{equation}
are {\em time-independent}. This is because the ratio
\begin{equation}
\label{ratio}
\tan \alpha=\frac{b_x(t)}{b_z(t)}
\end{equation}
remains constant at all times. In terms of $\alpha$,
the eigenvectors are
\begin{equation}
\label{eigvecs}
\psi_1= \left(\!\begin{array}{c}
\cos(\alpha/2)\\
\sin(\alpha/2)
\end{array}\!\right),\quad
\psi_2= \left(\!\begin{array}{c}
-\sin(\alpha/2)\\
\cos(\alpha/2)
\end{array}\!\right).
\end{equation}
The initial condition, $\langle \sigma_z(0)\rangle =1$,
corresponds to the following form of the wavefunction at $t=0$
\begin{equation}
\label{eigvecs}
\psi(0)= \left(\!\begin{array}{c}
1\\
0
\end{array}\!\right)= \cos\left(\frac \alpha 2\right ) \psi_1
-\sin\left(\frac \alpha 2\right ) \psi_2.
\end{equation}
The time evolution of this wavefunction is given by
\begin{equation}
\label{teigvecs} \psi(t)= \cos \! \left(\frac \alpha 2\right )
\psi_1 \exp\!\! \left[-\frac i2 \phi(t)\right] -\sin\! \left(\frac
\alpha 2\right ) \psi_2 \exp\!\! \left[\frac i2 \phi(t)\right]\! ,
\end{equation}
where the phase, $\phi(t)$, is defined as
\begin{equation}
\label{phi}
\phi(t)= \frac1{\cos\alpha}\int_0^tdt'b_z(t').
\end{equation}
Using Eq. (\ref{teigvecs}), we find the quantum-mechanical average
of the operator $\sigma_z$
\begin{equation}
\label{qmSz} \langle \psi(t)|\sigma_z|\psi(t)\rangle =
\cos^2\alpha + \sin^2\alpha \cos \phi(t).
\end{equation}
The remaining task is to perform the averaging over the noise
realizations. Note that only $\cos \phi(t)$ has to be averaged.
The details of this averaging are given in Appendix \ref{AppCorr}.
The final result reads
\begin{eqnarray}
\label{nobias} \langle \sigma_z(t)\rangle= \frac{b_z^2}{b^2}
+\frac{b_x^2}{b^2} e^{-t/\tau}\left[\cosh\left(
\frac t\tau \sqrt{1-b^2\tau^2}\right) \right .&&\\
\left . +\frac1{\sqrt{1-b^2\tau^2}} \sinh\left(
\frac t\tau \sqrt{1-b^2\tau^2}\right) \right],&& \nonumber
\end{eqnarray}
where $b$ is the magnitude of the full field
\begin{equation}
\label{b}
b=\big(b_x^2+b_z^2\big)^{1/2}.
\end{equation}
In the limit of a fast telegraph noise $b\tau \ll 1$
the result Eq. (\ref{nobias}) simplifies to

\begin{equation}
\label{fast}
\langle \sigma_z(t)\rangle= \frac{b_z^2}{b^2}
+\frac{b_x^2}{b^2}\exp\Bigg(-\frac{b^2\tau t}{2}    \Bigg).
\end{equation}
In the next subsection we compare this result with the NIBA prediction  Eq. (\ref{SzsolNIBA}).

\subsection{Comparison of exact and NIBA results}

\begin{figure}[t]
\centerline{\includegraphics[width=90mm,angle=0,clip]{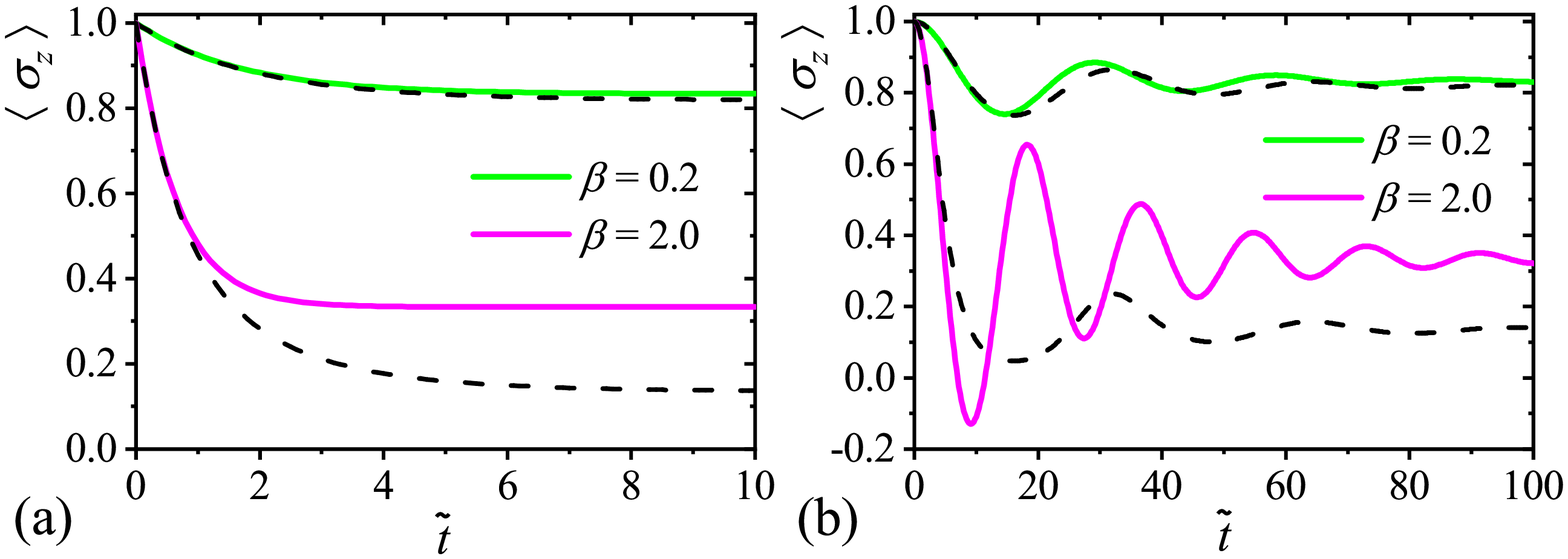}}
\caption{(Color online) Comparison of the exact and NIBA solutions
for $\langle\sigma_z(t)\rangle$.  (a) $\langle\sigma_z(t)\rangle$
is plotted versus the dimensionless time from Eq. (\ref{dimless})
(solid lines) and Eq. (\ref{xxx}) (dotted lines) for two values of
$\beta$ in the regime of fast noise $b_z\tau=0.1$. (b) The same
expressions are plotted for slow noise $b_z\tau =5$. The exact and
the NIBA solutions coincide at small times. The difference at long
times develops when $\beta$ is large.} \label{comparison}
\end{figure}

Expression Eq. (\ref{nobias}) is our central result. It describes
the evolution of $\langle \sigma_z(t)\rangle$ for arbitrary
relation between $b_x$, $b_z$, and $\tau^{-1}$. Our main message
that equal distribution of a particle between the sites is slowed
down at finite $b_z$, i.e. when the asymmetry between the sites
fluctuates in-phase with tunneling amplitude, is captured by this
expression. Indeed, for a fast noise, both $b_x\tau$ and $b_z\tau$
are small. At $b_z=0$, Eq.  (\ref{nobias}) yields $\langle
\sigma_z(t)\rangle =\exp\left(-b_x^2\tau t  \right)$, which
corresponds to the conventional motional narrowing.
\cite{Anderson1953, Kubo1, Anderson1962} It is seen from Eq.
(\ref{nobias}) that, even for $b_z\ll b_x$, the imbalance of the
population of sites does not decay down to zero at long times but
rather saturates at a finite value $\frac{b_z^2}{b^2}$. The
characteristic decay time becomes
$(b^2\tau)^{-1}$, i.e. it gets shorter. To analyze the dynamics of
$\langle \sigma_z(t)\rangle$ quantitatively we introduce the
dimensionless parameters
\begin{equation}
\label{parameters}
{\tilde t}=b_z^2\tau t,~~\beta=\frac{b_x^2}{b_z^2}=\tan^2\alpha.
\end{equation}
Then Eq. (\ref{nobias}) assumes the form
\begin{widetext}
\begin{eqnarray}
\label{dimless} \langle \sigma_z(t)\rangle= \frac 1{1+\beta} +
\frac \beta{2(1+\beta)} \left\{
\frac{\sqrt{1-(b_z\tau)^2(1+\beta)}+1}{\sqrt{1-(b_z\tau)^2(1+\beta)}}
\exp\left[-  \frac{\tilde{t}}{(b_z\tau)^2}\left(
1-\sqrt{1-(b_z\tau)^2(1+\beta)}\right)
\right] \right . &&\nonumber \\
\left . +\frac{\sqrt{1-(b_z\tau)^2(1+\beta)}-1}
{\sqrt{1-(b_z\tau)^2(1+\beta)}}\exp\left[-
\frac{\tilde{t}}{(b_z\tau)^2}\left(
1+\sqrt{1-(b_z\tau)^2(1+\beta)}\right) \right] \right \}&&.
\end{eqnarray}
We will contrast this expression to the NIBA result
Eq. (\ref{SzsolNIBA}),
which, in variables Eq. (\ref{parameters}), takes the form
\begin{eqnarray}
\label{xxx} \langle \sigma_z(t)\rangle_{\scriptscriptstyle \text
{NIBA}}= \exp \left\{-\beta -\frac{\beta
(b_z\tau)^2}{2\sqrt{1-(b_z\tau)^2}} \left ( \frac
1{1-\sqrt{1-(b_z\tau)^2}}\exp\left[-
\frac{\tilde{t}}{(b_z\tau)^2}\left( 1+\sqrt{1-(b_z\tau)^2}\right)
\right] \right . \right .&& \nonumber\\
\left . \left . - \frac 1{1+\sqrt{1-(b_z\tau)^2}}\exp\left[-  \frac{\tilde{t}}{(b_z\tau)^2}\left(
1-\sqrt{1+(b_z\tau)^2}\right)
\right]\right) \right \}.&&
\end{eqnarray}
\end{widetext}
Both expressions are plotted in Fig. \ref{comparison} for
different values of parameters $\beta$ and different noise
``strengths" $b_z\tau$. We see that NIBA reproduces the exact
result at small $\beta$, i.e. at $b_x\ll b_z$. This could be anticipated from the
analytical expression Eq. (\ref{xxx})
since it saturates at
$\exp(-\beta)$, while the exact saturation value is
$\langle\sigma_z(\infty)\rangle =\frac{1}{1+\beta}$. At small
$\beta$ the net change of  $\langle\sigma_z\rangle$ from $t=0$ and
$t=\infty$ is small. This justifies taking
$\langle\sigma_z(t)\rangle$ out of the integrand in the NIBA
equation Eq. (\ref{SzNIBA}). It is worth noting that both
solutions oscillate at large $b_z\tau$. Emergence of these weakly
decaying oscillations is not obvious a priori since they are
noise-induced. The oscillations become possible because, for
telegraph noise, the full field switches between two fixed values.

The fact that $\langle \sigma_z(t)\rangle$ can be calculated
exactly for the telegraph-noise fluctuations of $b_x(t)$, $b_z(t)$
is long known in the literature on stochastic
dynamics.\cite{vanKampen} In particular, in the chapter of the
book Ref. \onlinecite{vanKampen} on stochastic differential
equations, the following problem is considered. Suppose that the
frequency of an oscillator takes only two values with equal
probabilities. Then the noise-averaged dynamics of the oscillator,
derived from the Fokker-Planck description, has the form closely
resembling Eq. (\ref{nobias}). In fact, our result Eq.
(\ref{nobias}) reduces to the result\cite{vanKampen} for
two-frequency oscillator problem if we set $b_z(t)=0$. This is the
extreme ``non-NIBA" limit. Presence of both   $b_x(t)$ and
$b_z(t)$ fluctuating in-phase can be viewed as a vector version of
the two-frequency oscillator problem. We have demonstrated that
incorporating even small $b_z(t)$ leads to a dramatic physical
consequence, namely, saturation of $\langle \sigma_z(t)\rangle$ instead of
decay.

\section{Finite bias}

\subsection{Spin dynamics within NIBA}

\begin{figure}[t]
\centerline{\includegraphics[width=90mm,angle=0,clip]{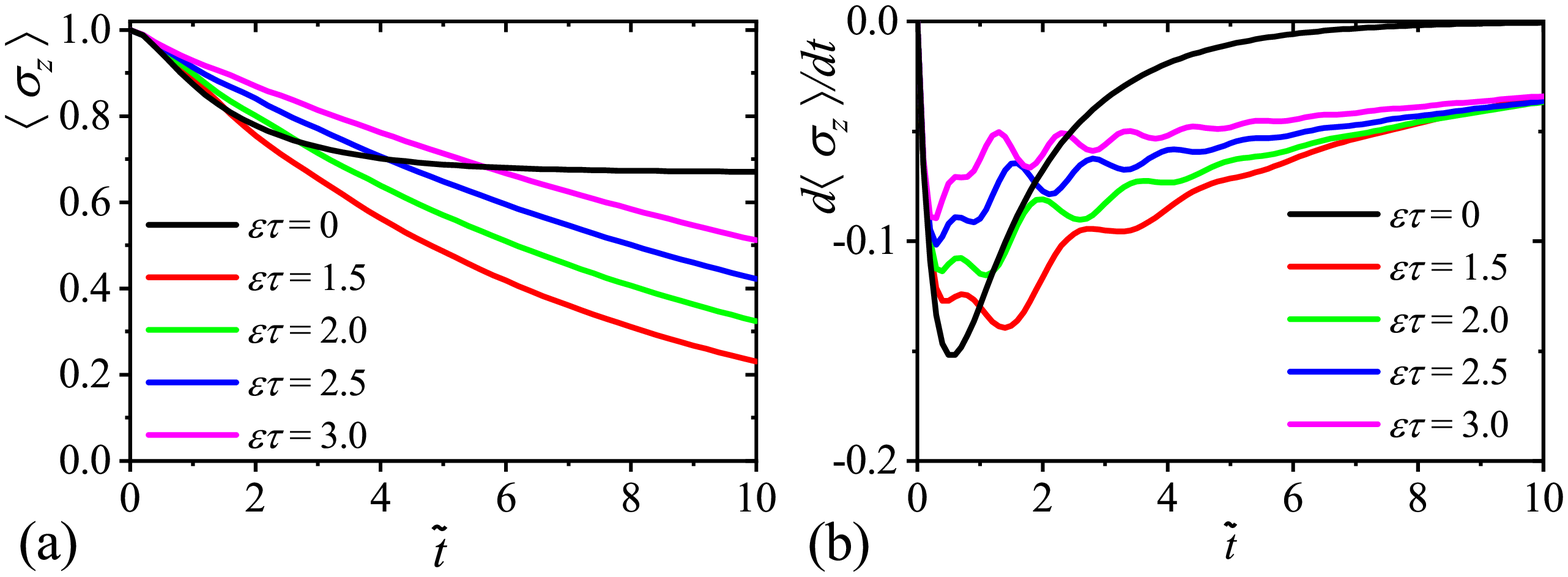}}
\caption{(Color online) Plotted are the time dependencies of
$\langle \sigma_z(t)\rangle$ (a) and $d\langle
\sigma_z(t)\rangle/dt$ (b) obtained within NIBA, for different
values of bias $\varepsilon\tau =0, 1.5, 2, 2.5$, and $3$. In
plotting these curves from Eq. (\ref{SbiasNIBA}) the parameter
values are set to $b_z\tau=0.5$, $\beta= 0.2$.} \label{oscNIBA}
\end{figure}

In the presence of a finite bias, $\varepsilon$, which adds to the
random field $b_z(t)$, the NIBA equation Eq. (\ref{SzNIBA})
assumes the form
\begin{eqnarray}
\label{SzNIBAbias}
&&\frac{d \langle \sigma_z \rangle}{d t} =\\
&&- \! \int
\limits_0^t \!dt_1\left \langle \! b_x(t) b_x(t_1)
\cos\left[\varepsilon(t-t_1)+\int\limits_{t_1}^t \!dt_2
b_z(t_2) \right ]\right \rangle \langle \sigma_z(t_1) \rangle. \nonumber
\end{eqnarray}
The solution of Eq. (\ref {SzNIBAbias}) is a straightforward generalization of
Eq. (\ref{SzsolNIBA})
\begin{eqnarray}
\label{SzsolNIBAbias}
&&\langle \sigma_z(t)\rangle =\exp\Big \{\!-\int\limits_0^tdt_1\int\limits_0^{t_1}dt_2
\cos\left[\varepsilon(t_1-t_2)\right]K(t_1-t_2)\Big\} \nonumber\\
&&=\exp\Bigg[-\int\limits_0^t dt_1(t-t_1)\cos\varepsilon (t-t_1) K(t_1)  \Bigg].
\end{eqnarray}
With correlator Eq. (\ref{KT}), the integration can be performed analytically
\begin{widetext}
\begin{equation}
\label{SbiasNIBA} \langle \sigma_z(t)\rangle_{\scriptscriptstyle
\text {NIBA}} = \exp\big \{ -\kappa t -F_0 -F_c(t) \cos
\varepsilon t -F_s(t)\sin \varepsilon t \big \},
\end{equation}
where
\begin{eqnarray}
\label{longest}
&&\kappa=\frac{b_x^2\tau\varepsilon^2(\tau_s^2-\tau_f^2)}
{2\sqrt{1-(b_z\tau)^2}(1+\varepsilon^2\tau_f^2)(1+\varepsilon^2\tau_s^2)},
\qquad F_0 = -\frac{b_x^2\tau}{2\sqrt{1-(b_z\tau)^2}} \left [
\frac{\tau_f(1-\varepsilon^2\tau_f^2)}{(1+\varepsilon^2\tau_f^2)^2}
-\frac{\tau_s(1-\varepsilon^2\tau_s^2)}{(1+\varepsilon^2\tau_s^2)^2}
\right ],
\nonumber  \\
&&F_c= \frac{b_x^2\tau}{2\sqrt{1-(b_z\tau)^2}} \left [
\frac{\tau_f(1-\varepsilon^2\tau_f^2)}
{(1+\varepsilon^2\tau_f^2)^2}\exp \left ( -\frac t{\tau_f} \right)
-\frac{\tau_s(1-\varepsilon^2\tau_s^2)}{(1+\varepsilon^2\tau_s^2)^2}\exp
\left ( -\frac t{\tau_s} \right)
 \right ], \nonumber \\
&&F_s= -\frac{b_x^2\tau\varepsilon}{\sqrt{1-(b_z\tau)^2}} \left [
\frac{\tau_f^2} {(1+\varepsilon^2\tau_f^2)^2}\exp \left ( -\frac
t{\tau_f} \right)
-\frac{\tau_s^2}{(1+\varepsilon^2\tau_s^2)^2}\exp \left ( -\frac
t{\tau_s} \right)
 \right ].
\end{eqnarray}
\end{widetext}
First we note that, at arbitrary nonzero bias, $\langle
\sigma_z(t)\rangle \to 0$ instead of saturation at long times. The
decay of $\langle \sigma_z(t)\rangle$ is governed by the factor
$\exp(-\kappa t)$ in Eq. (\ref{SbiasNIBA}), in which $\kappa$ is
proportional to $\varepsilon^2$. Presence of the terms $F_c$ and
$F_s$ in the exponent of Eq. (\ref{SbiasNIBA}) suggests
oscillatory behavior of $\langle \sigma_z(t)\rangle$ with the
frequency $\varepsilon$ at intermediate times. On the other hand,
the prefactors $F_c$ and $F_s$ decay at long times. Thus it is not
clear a priori whether these oscillations can be resolved. In Fig.
\ref{oscNIBA} we present numerical curves plotted from Eq.
(\ref{SbiasNIBA}) for different biases. We realize that
oscillations are not developed in $\langle \sigma_z(t)\rangle$ but
are
resolved in the derivatives, $d\langle \sigma_z(t)\rangle/dt$.
It is an interesting question whether these oscillations survive
beyond NIBA. We address this question in the next section.

\subsection{Beyond NIBA: small bias limit}

Exact solution Eq. (\ref{qmSz}) applies not only to the telegraph
noise but to arbitrary $b_x(t)$ and $b_z(t)$ as long as the ratio
$\frac{b_x}{b_z}=\tan\alpha$ remains constant. We will derive the
evolution $\langle \sigma_z(t)\rangle$ assuming that the bias,
$\varepsilon$, is much smaller than $b$. We start from the system
of equations for ``up" and ``down" amplitudes of spin
\begin{eqnarray}
\label{system}
&& i\frac{dc_1}{dt}= \frac 12\Bigl[b_z(t)+\varepsilon\Bigr]c_1
+ \frac 12b_x(t)c_2, \nonumber \\
&& i\frac{dc_2}{dt}=-\frac 12 \Bigl[b_z(t) +\varepsilon\Bigr]c_2 +
\frac 12 b_x(t)c_1.
\end{eqnarray}
As in the case of zero bias, we introduce the linear combinations
\begin{eqnarray}
\label{variables} &&c_1=u_1\cos\left(\frac{\alpha}{2}
\right)-u_2\sin\left(\frac{\alpha}{2} \right), \nonumber \\
&&c_2=u_1\sin\left(\frac{\alpha}{2} \right)
+u_2\cos\left(\frac{\alpha}{2} \right).
\end{eqnarray}
The system of equations for the new variables $u_1$ and $u_2$
reads
\begin{eqnarray}
\label{SYSTEM}
&&i\frac{du_1}{dt}=\frac{b_z+\varepsilon \cos^2\alpha}{2\cos \alpha}u_1
-\frac 12 \varepsilon \sin\alpha~u_2, \nonumber\\
&&i\frac{du_2}{dt}=-\frac{b_z +\varepsilon \cos^2\alpha}{2\cos
\alpha}u_2 -\frac 12\varepsilon \sin\alpha~u_1.
\end{eqnarray}
It is easy to see that at zero bias, $\varepsilon=0$, the system
Eq.~(\ref{SYSTEM}) gets decoupled. Using the initial conditions
\begin{equation}
\label{initial} u_1(0)=\cos\Bigl(\frac{\alpha}{2}\Bigr), \quad
u_2(0)=-\sin\Bigl(\frac{\alpha}{2}\Bigr),
\end{equation}
the exact result Eq. (\ref{qmSz}) can be reproduced.

At finite $\varepsilon$, in Eq. (\ref{SYSTEM}) we make the
substitution
\begin{eqnarray}
\label{substitution}
&&u_1(t)=v_1(t)\exp\left[-\frac i2\Phi(t) \right], \nonumber \\
&&u_2(t)=v_2(t)\exp\left[\frac i2\Phi(t) \right],
\end{eqnarray}
were we have introduced the short-hand notation
\begin{equation}
\label{beta} \Phi(t)=\frac 1{\cos\alpha} \int_0^t dt' \left[
b_z(t')+\varepsilon \cos^2\alpha   \right].
\end{equation}
For $\varepsilon=0$ we have $\Phi(t)=\phi(t)$, where $\phi$ is
defined by Eq. (\ref{phi}). The substitution Eq.
(\ref{substitution}) yields the following system of coupled
equations for the variables $v_1(t)$, $v_2(t)$
\begin{eqnarray}
\label{coupled} &&i\frac{dv_1}{dt}=-\frac 12\varepsilon
v_2\sin\alpha
\exp\bigl[i\Phi(t) \bigr],\\
&&i\frac{dv_2}{dt}=-\frac 12\varepsilon v_1\sin\alpha
\exp\bigl[-i\Phi(t) \bigr].
\end{eqnarray}
Substituting the second equation into the first,
we arrive to the closed integral-differential  equation for $v_1(t)$
\begin{eqnarray}
\label{closed}
\frac{dv_1}{dt}=-i\frac\varepsilon 2\sin\alpha~v_2(0)\exp\bigl[i\Phi(t) \bigr]&& \nonumber \\
-\frac{\varepsilon^2}4\sin^2\alpha\int\limits_0^tdt'v_1(t')\exp\Bigl[i\bigl(\Phi(t)
- \Phi(t')\bigr) \Bigr].&&
\end{eqnarray}
This equation
applies for {\em arbitrary bias}. At this point we note that,
since the  right-hand side is proportional to $\varepsilon$, for
small $\varepsilon$ the derivative $\frac{dv_1}{dt}$ is small.
Thus, the function $v_1(t)$ changes slowly with time. This allows
one to pull $v_1(t)$  out of the  integrand. Then Eq.
(\ref{closed}) turns into a first-order differential equation
which can be readily solved yielding
\begin{eqnarray}
\label{solved} v_1(t)=v_1(0)
\exp\Bigl[-\frac{\varepsilon^2}4\sin^2\alpha~
\int\limits_0^t dt'G(t') \Bigr]&& \nonumber\\
+ i\frac\varepsilon 2\sin\alpha~v_2(0)\!\! \int\limits_0^t \!\!
dt'\! \exp\Bigl[i\Phi(t')
-\frac{\varepsilon^2}4\sin^2\alpha\bigl(G(t)-G(t')\bigr)
\Bigr],&&\nonumber \\
\end{eqnarray}
%
where the function $G(t)$ is defined as
\begin{equation}
G(t)=\int\limits_0^tdt' \exp\Bigl[i\bigl(\Phi(t) - \Phi(t')\bigr)
\Bigr].
\end{equation}
Corresponding expression for $v_2(t)$ follows from Eq. (\ref{solved}) upon replacement
$v_1(0)\rightarrow v_2(0)$, $v_2(0)\rightarrow v_1(0)$, and $\Phi(t) \rightarrow -\Phi(t)$.

The sought quantity, $\sigma_z(t)$,  is expressed via the
functions $v_1(t)$ and $v_2(t)$ as follows
\begin{eqnarray}
\label{AVERAGE}
\sigma_z(t)=\Bigl(|v_1|^2-|v_2|^2\Bigr)\cos\alpha&& \nonumber\\
-\sin\alpha\Bigl(v_1v_2^* \exp\exp\bigl[-i\Phi(t)
\bigr]+v_1^*v_2\exp\bigl[i\Phi(t)
\bigr]       \Bigr).&& \nonumber \\
\end{eqnarray}
A crucial step in performing averaging in Eq. (\ref{AVERAGE}) is
that the second term in the expression for $v_1(t)$ proportional
to $\varepsilon$ is small compared to the first term. The result
of averaging over realizations of the telegraph noise reads
\begin{eqnarray}
\label{weakbias} \langle \sigma_z(t)\rangle \! \! &=& \! \!\nonumber\\
\Biggl[\frac{b_z^2}{b^2} &+& \! \! \frac{b_x^2}{b^2} \exp\!
\left(\! -\frac{b^2\tau}2 t\right)\! \cos\! \left ( \varepsilon
\frac{b_z} b t \right)\! \Biggr]\!
\exp\! \left(\! -2\varepsilon^2\frac{b_x^2}{b^4\tau}t\right)\! .\nonumber\\
\end{eqnarray}
The origin of the oscillating factor $\cos \left ( \varepsilon
\frac{b_z} b t \right)$ is the term $\sim\varepsilon$ in $\Phi(t)$
defined by Eq. (\ref{beta}). The common exponential factor
originates from averaging of $\exp \Bigl[ -\frac{\varepsilon^2} 4
\sin^2\alpha~ \int\limits_0^t dt'G(t') \Bigr]$.

We can now compare the result Eq. (\ref{weakbias}) with the NIBA result Eq. (\ref{SbiasNIBA}).
The decrement of  $\langle \sigma_z(t)\rangle_{\scriptscriptstyle
\text {NIBA}}$ in the limit $b_z\tau \ll 1$ can be cast in the form
\begin{equation}
\label{decrement}
\kappa=\frac{b_x^2\tau \varepsilon^2\tau_s^2}{2\left( 1+\varepsilon^2\tau_s^2 \right)}=\frac{\varepsilon^2\tau_sb_x^2}
{ \left( 1+\varepsilon^2\tau_s^2 \right)b_z^2         }.
\end{equation}
where $\tau_s\approx \frac{2}{b_z^2\tau}$ is defined by Eq. (\ref{taufs}).
In the limit of small bias considered above we  have $\varepsilon \tau_s \ll 1$.
Then the decrement Eq. (\ref{decrement}) reproduces the common exponential factor
in Eq. (\ref{weakbias}) under the condition $b_x \ll b_z$.
This is the same condition under which  NIBA applies at zero bias.
Equally, expanding the exponent in the  NIBA result with respect
to $F_c(t)\cos\varepsilon t$ and assuming $b_x\ll b_z$, we reproduce the oscillating part of
Eq. (\ref{weakbias}).

Overall, the small-bias regime is quantified by
the condition $\varepsilon \tau_s \ll 1$. This justifies
the reduction of Eq. (\ref{closed}) to the first-order
differential equation. Also the second term in Eq. (\ref{solved})
is of the order of $\varepsilon\tau_s$.

\section{Discussion}

1. Physical situation considered in the present paper
corresponds to the noise created by a fluctuator, see e.g.
Ref. \onlinecite{Galperin}, rather than the noise created
by the continuum of harmonic oscillators commonly considered
in the literature on quantum dissipation.

2. Our central conclusion is that, when the  components $b_x(t)$
and $b_z(t)$ are fully correlated, the average $\sigma_z(t)$
saturates at long times.  At small $b_z \ll b_x$ it saturates at
small but finite value. This statement actually applies for
arbitrary noise with $b_z(t)$ and $b_x(t)$ having the same time
dependence. Indeed, Eq. (\ref{qmSz}) ``knows" about the noise only
via a random phase $\phi(t)$, while non-vanishing term
$\cos^2\alpha= b_z^2/b^2$ is time-independent. The saturation
value, $\langle\sigma_z(\infty)\rangle$, is captured by NIBA
correctly when $b_x \ll b_z$. In the language of the two-site model
this statement implies that for NIBA to apply the
 magnitude of the fluctuating tunneling
amplitude should be much smaller than the magnitude of the
fluctuating level splitting.
Note that anomalously strong sensitivity of the spin dynamics to the finite bias
is long known in the field of quantum dissipation. \cite{1,2,3,4,5}

3. NIBA result Eq. (\ref{SbiasNIBA}) contains two times, $\tau_s$ and $\tau_f$.
Exact result also contains two characteristic times
but instead of the combination $\sqrt{1-b_z^2\tau^2}$
these times contain $\sqrt{1-(b_z^2+b_x^2)\tau^2}$.
NIBA applies in the limit $b_x\ll b_z$ when these times
are close to each other.

4. Unlike Ref. \onlinecite {Imambekov}, $\langle \sigma_z(t)\rangle$ saturates at long
times. The meaning of $\langle \sigma_z(t)\rangle$ is the average
over the realizations of the telegraph noise. In fact, the
derivation of the result  Eq. (\ref{qmSz}) can be modified to find
the variance
\begin{equation}
\label{var}
{\text Var}(\sigma_z) = \sin^4\alpha \big[
\langle \cos^2\phi(t) \rangle -\langle\cos\phi(t)\rangle^2
\big ].
\end{equation}
Analytical form of $\langle\cos\phi(t)\rangle$ was found above. To
find $\langle\cos^2\phi(t)\rangle$ one can use the same expression
with $b$ replaced by $2b$. This follows from the relation
\begin{equation}
\label{2phi}
\cos\big [2\phi(t)\big] = \cos\left[ \frac2{\cos\alpha}\int_0^tdt'b_z(t')\right].
\end{equation}
In the long-time limit the variance saturates together with average
\begin{equation}
\label{varinfty} {\text Var}\big[\sigma_z(t)\big]\Big |_{t\to
\infty}  = \frac 12\sin^4\alpha =\frac{b_x^4}{2b^4}.
\end{equation}
Note that the fluctuations of $\sigma_z(\infty)$ are
much smaller than the average when $b_x \ll b_z$, i.e.
in the same domain where NIBA applies. In the opposite
limit, $b_x\gg b_z$, fluctuations of $\sigma_z(t)$  from realization to
realization are strong.

%

\section{Acknowledgements}
This work was supported by the Department of Energy,
Office of Basic Energy Sciences, Grant No. DE-FG02-06ER46313.

\appendix

\section{}

\label{AppCorr}


In this Appendix we evaluate the average, $\langle
\cos\phi(t)\rangle$, where the phase $\phi(t)$ is defined by Eq.
(\ref{phi}). In the course of the noise, the magnetic field
switches from $-b$ to $b$ at random time moments. The durations of
intervals between the successive switchings are
Poisson-distributed as $\frac{1}{\tau}\exp\left(-\frac{t}{\tau}
\right)$. Thus, averaging over the noise realizations reduces to
averaging over these intervals
\begin{eqnarray}\label{avcos}
\langle \cos\phi(t)\rangle = \text{ Re} \sum_{n=0}^\infty
\int\limits_0^\infty\! \frac{dt_1}\tau\, e^{-\frac{t_1}\tau}
\cdots\! \int\limits_0^\infty\! \frac{dt_{n+1}}{\tau}\,
e^{-\frac{t_{n+1}}\tau}
\nonumber\\
\times e^{ib(t_1-t_2+\cdots (-1)^nt_{n+1} )}\left[\theta\big(\!
{\scriptstyle t-\sum\limits_{k=1}^nt_k }\!\big) -\theta\bigr(\!
{\scriptstyle t-\sum\limits_{k=1}^{n+1}t_k }\!\bigl) \right]\!.&&
\end{eqnarray}
Here $n$ is the number of field flips during the time $t$. This is
ensured by the difference of the $\theta$-functions which imposes
the condition $\sum_{k=1}^nt_k < t < \sum_{k=1}^{n+1}t_k$.

Taking the integral over $t_{n+1}$ by parts yields
\begin{eqnarray}
\label{avcos1} \langle \cos\phi(t)\rangle = \text{ Re}
\sum_{n=0}^\infty \int\limits_0^\infty\! \frac{dt_1}\tau \cdots\!
\int\limits_0^\infty\! \frac{dt_n}{\tau} \int\limits_0^\infty\!
\frac{dt_{n+1}}{\tau}\,
e^{-\sum\limits_{j=1}^{n+1}\frac{t_j}\tau}&&
\nonumber\\
\times e^{ib(t_1-t_2+\cdots (-1)^nt_{n+1}
)}\delta\bigr({\scriptstyle t-\sum\limits_{k=1}^{n+1}t_k
}\bigl).\quad &&
\end{eqnarray}
The individual integrals over $t_i$ can be taken upon using the
integral representation, $\delta(z) =\int_{-\infty}^\infty
\frac{ds}{2\pi}e^{isz}$. This leads to
\begin{equation}\label{avcos2}
\langle \cos\phi(t)\rangle = \text{ Re} \sum_{k=1}^\infty
\int\limits_{-\infty}^\infty \! \frac{ds}{2\pi}\, e^{ist}\frac
{2\tau+ i(s+b)\tau^2} {[(1 +is\tau)^2 +b^2\tau^2]^k}.
\end{equation}
We next take the sum in Eq. (\ref{avcos2}) and symmetrize the
result with respect to the sign of $b$ at $t=0$. This yields
\begin{equation}\label{avcos3}
\langle \cos\phi(t)\rangle = \text{ Re} \int
\limits_{-\infty}^\infty \! \frac{ds}{2\pi}\, e^{ist}\frac {2\tau
+is\tau^2} {(1 +is\tau)^2 +b^2\tau^2-1}.
\end{equation}
The latter integral is calculated by adding up the contributions
of two poles, located at $s=\frac{i}{\tau} \pm
\left(b^2-\frac{1}{\tau^2}\right)^{1/2}$. For $b\tau<1$, we find
\begin{eqnarray}
\label{avcos1}
&&\langle \cos\phi(t)\rangle = \exp{\left(-\frac t\tau\right)}\\
&&\times \left\{\!\cosh\! \left[ \frac t\tau
\left(1-b^2\tau^2\right)^{1/2} \right] +\frac{\sinh\!\left[ \frac
t\tau \left(1-b^2\tau^2\right)^{1/2}
\right]}{\left(1-b^2\tau^2\right)^{1/2}} \right\}.\nonumber
\end{eqnarray}
Substituting Eq. (\ref{avcos1}) into Eq. (\ref{qmSz}) leads to our
main result Eq. (\ref{nobias}).

To find the average defining the kernel $K(T)$ through
Eq.~(\ref{KT}) we note the relation,
\begin{equation}
\label{ddrel} K(t_2-t_1)=\Bigg(\frac{b_x^2}{b^2}\Bigg)
\frac{\partial^2} {\partial t_1 \partial t_2} \langle
\cos\phi(t_2-t_1)\rangle.
\end{equation}
This relation yields
\begin{eqnarray}
\label{KTder}
&&K(T) = b_x^2\exp{\left(-\frac T\tau\right)}\\
&&\times \left\{\!\cosh\! \left[ \frac T\tau
\left(1-b^2\tau^2\right)^{1/2} \right] -\frac{\sinh\!\left[ \frac
T\tau \left(1-b^2\tau^2\right)^{1/2}
\right]}{\left(1-b^2\tau^2\right)^{1/2}} \right\}.\nonumber
\end{eqnarray}
Note that two terms in $\{..\}$  in Eq.  (\ref{avcos1}) add up,
while in Eq. (\ref{KTder}) they subtract.


\begin{thebibliography}{30}





\bibitem{1} A. O. Caldeira and A. J. Leggett, ``Influence of Dissipation on Quantum Tunneling
in Macroscopic Systems," Phys. Rev. Lett. {\bf 46}, 211 (1981).

\bibitem{2} A. J. Bray and M. A. Moore,``
Influence of Dissipation on Quantum Coherence,"
Phys. Rev. Lett. {\bf 49}, 1545 (1982).

\bibitem{3} H. Grabert and U. Weiss,
``Quantum Tunneling Rates for Asymmetric Double-Well Systems with Ohmic Dissipation,"
Phys. Rev. Lett. {\bf 54}, 1605 (1985).

\bibitem{4}
A. J. Leggett, S. Chakravarty, A. T. Dorsey, M. P. A. Fisher,
A. Garg, and W. Zwerger,
``Dynamics of the dissipative two-state system,"
Rev. Mod. Phys. {\bf 59}, 1 (1987).


\bibitem{5} M. Grifoni and P. H{\"a}nngi,
``Driven quantum tunneling,"
Phys. Rep. {\bf 304}, 229 (1998).

\bibitem{6} K. Le Hur, ``Quantum Phase Transitions in Spin-Boson Systems: Dissipation and Light Phenomena," in
{\em Understanding Quantum Phase Transitions}, edited by Lincoln D. Carr (Taylor and Francis, Boca Raton, 2010).


\bibitem{7}  U. Weiss, {\em Quantum Dissipative Systems,} 4th ed. (World Scientific, Singapore, 2012).

\bibitem{8}I. de Vega and D. Alonso,
``Dynamics of non-Markovian open quantum systems,"
Rev. Mod. Phys. {\bf 89}, 015001 (2017).


\bibitem{9} L. M. Cangemi, V. Cataudella, M. Sassetti, G. De Filippis,
``Dissipative dynamics of a driven qubit: interplay between non-adiabatic dynamics and noise effects from weak to strong coupling regime,"
Phys. Rev. B {\bf 100}, 014301 (2019).



\bibitem{C1} L. M. Cangemi, G. Passarelli, V. Cataudella, P. Lucignano, and G. De Filippis,
``Beyond the Born-Markov approximation: Dissipative dynamics of a single qubit,"
Phys. Rev. B {\bf 98}, 184306 (2018).

\bibitem{C2} M. Wiedmann, J. T. Stockburger, and J. Ankerhold,
``Time-correlated blip dynamics of open quantum systems,"
Phys. Rev. A {\bf 94}, 052137 (2016).

\bibitem{C3} S. Bera, H. U. Baranger, and S. Florens,
``Dynamics of a qubit in a high-impedance transmission line from a bath perspective,"
Phys. Rev. A {\bf 93}, 033847 (2016).

\bibitem{C4} H. Shapourian,
``Dynamical renormalization-group approach to the spin-boson model,"
Phys. Rev. A {\bf 93}, 032119 (2016).


\bibitem{C5} P. P. Orth, A. Imambekov, and K. Le Hur,
``Nonperturbative stochastic method for driven spin-boson model,"
Phys. Rev. B {\bf 87}, 014305 (2013).

\bibitem{C6}
F. Nesi, E. Paladino, M. Thorwart, and M. Grifoni,
``Spin-boson dynamics beyond conventional perturbation theories,"
Phys. Rev. B {\bf 76}, 155323 (2007).


\bibitem{C7} M. Thoss, H. Wang, and W. H. Miller,
``Self-consistent hybrid approach for
complex systems: Application
to the spin-boson model with Debye spectral density,"
J. Chem. Phys. {\bf 115}, 2991 (2001).


\bibitem{C8} C. H. Mak and R. Egger,
``Quantum Monte Carlo study of tunneling diffusion in a dissipative multistate system," Phys. Rev. E {\bf 49}, 1997 (1994).



\bibitem{Imambekov} G. B. Lesovik, A. V. Lebedev,
and A. O. Imambekov, ``Dynamics of Two-Level
System Interacting with Random Classical Field,"
JETP Lett. {\bf 75} 474, (2002).

\bibitem{Dekker1987}
This reformulation of the problem was first proposed by H. Dekker, ``Noninteracting-blip
approximation for a two-level system coupled to a
heat bath," Phys. Rev. A {\bf 35}, 1436 (1987).

\bibitem{we}
V. V. Mkhitaryan, C. Boehme, J. M. Lupton, and M. E. Raikh,
``Two-photon absorption in a two-level system enabled by noise,"
Phys. Rev B {\bf 100}, 214205 (2019).

\bibitem{Anderson1953}
P. W. Anderson and P. R. Weiss,
``Exchange Narrowing in Paramagnetic Resonance,"
Rev. Mod. Phys. {\bf 25}, 269 (1953).

\bibitem{Kubo1}
R. Kubo, ``Note on the Stochastic Theory of
Resonance Absorption," J. Phys. Soc. Jpn.
{\bf 9}, 935 (1954).

\bibitem{Anderson1962}
J. R. Klauder and P. W. Anderson,
``Spectral Diffusion Decay in Spin Resonance Experiments,"
Phys. Rev. {\bf 125}, (1962).

\bibitem{vanKampen} N. van Kampen,
{\em Stochastic Processes in Physics and Chemistry} (North-Holland, 1981).

\bibitem{Galperin}
E. Paladino, Y. M. Galperin, G. Falci, and B. L. Altshuler,
``$1/f$ noise: Implications for solid-state quantum information,"
Rev. Mod. Phys. {\bf 86}, 361 (2014).
\end{thebibliography}
\end{document}